\documentstyle[11pt]{article}

\def\dalemb#1#2{{\vbox{\hrule height .#2pt
        \hbox{\vrule width.#2pt height#1pt \kern#1pt
                \vrule width.#2pt}
        \hrule height.#2pt}}}
\def\square{\mathord{\dalemb{6.8}{7}\hbox{\hskip1pt}}}

\def\td{\tilde}
\def\wtd{\widetilde}

\let\a=\alpha    
    
  \let\n=\nu

\def\nn{\nonumber} \def\bd{\begin{document}} \def\ed{\end{document}}
\def\ds{\documentstyle} \let\fr=\frac \let\bl=\bigl \let\br=\bigr
\let\Br=\Bigr \let\Bl=\Bigl 
\let\bm=\bibitem
\let\na=\nabla
\let\pa=\partial \let\ov=\overline 
\newcommand{\be}{\begin{equation}} 
\newcommand{\ee}{\end{equation}} 
\def\ba{\begin{array}}
\def\ea{\end{array}}
\def\ft#1#2{{\textstyle{{\scriptstyle #1}\over {\scriptstyle #2}}}}
\def\fft#1#2{{#1 \over #2}}
\def\del{\partial}
\def\sst#1{{\scriptscriptstyle #1}}
\def\oneone{\rlap 1\mkern4mu{\rm l}}
\def\ie{{\it i.e.\ }}
\def\via{{\it via}}
\def\A{{\sst A}}
\def\B{{\sst B}}
\def\wdg{{\,\sst \wedge \,}}
\def\0{{\sst{(0)}}}
\def\1{{\sst{(1)}}}
\def\2{{\sst{(2)}}}
\def\3{{\sst{(3)}}}
\def\4{{\sst{(4)}}}
\def\5{{\sst{(5)}}}
\def\6{{\sst{(6)}}}
\def\7{{\sst{(7)}}}
\def\n{{\sst{(n)}}}
\newcommand{\ho}[1]{$\, ^{#1}$}
\newcommand{\hoch}[1]{$\, ^{#1}$}
\newcommand{\bea}{\begin{eqnarray}} 
\newcommand{\eea}{\end{eqnarray}} 
\newcommand{\ra}{\rightarrow}
\newcommand{\lra}{\longrightarrow}
\newcommand{\Lra}{\Leftrightarrow}
\newcommand{\ap}{\alpha^\prime}
\newcommand{\bp}{\tilde \beta^\prime}
\newcommand{\tr}{{\rm tr} }
\newcommand{\Tr}{{\rm Tr} } 
\newcommand{\NP}{Nucl. Phys. }
\newcommand{\tamphys}{\it Center for Theoretical Physics,
Texas A\&M University, College Station, Texas 77843}
\newcommand{\ens}{\it Laboratoire de Physique Th\'eorique de l'\'Ecole
Normale Sup\'erieure\hoch{2}\\
24 Rue Lhomond - 75231 Paris CEDEX 05}

\newcommand{\auth}{H. L\"u\hoch{\dagger} and
C.N. Pope\hoch{\ddagger1}}

\begin{document}
\begin{flushright}
\hfill{CTP TAMU-41/97}\\
\hfill{LPTENS-97/46}\\
\hfill{hep-th/9710155}\\
\hfill{Oct. 1997}\\
\end{flushright}

\vspace{15pt}

\begin{center}
{ \large {\bf Interacting Intersections}}

\vspace{20pt}
\auth

\vspace{15pt}

{\hoch{\dagger}\ens}

\vspace{10pt}
{\hoch{\ddagger}\tamphys}

\vspace{40pt}

\underline{ABSTRACT}
\end{center}

   Intersecting $p$-branes can be viewed as higher-dimensional 
interpretations of multi-charge extremal $p$-branes, where some of the
individual $p$-branes undergo diagonal dimensional
oxidation, while the others oxidise vertically.  Although the naive
vertical oxidation of a single $p$-brane gives a continuum of
$p$-branes, a more natural description arises if one considers a
periodic array of $p$-branes in the higher dimension, implying a dependence
on the compactification coordinates. This 
still reduces to the single lower-dimensional $p$-brane when viewed
at distances large compared with the period.  Applying the same logic
to the multi-charge solutions, we are led to consider more general
classes of intersecting $p$-brane solutions, again depending on the 
compactification coordinates, which turn out to be 
described by interacting functions rather than independent harmonic
functions. These new solutions also provide a more satisfactory
interpretation for the lower-dimensional multi-charge $p$-branes,
which otherwise appear to be nothing more than the improbable
coincidence of charge-centres of individual constituents with zero
binding energy.

{\vfill\leftline{}\vfill
\vskip	10pt
\footnoterule
{\footnotesize	\hoch{1} Research supported in part by DOE 
Grant DE-FG03-95ER40917 \vskip	-12pt} \vskip 14pt
{\footnotesize
        \hoch{2} Unit\'e Propre du Centre National de la Recherche
Scientifique, associ\'ee \`a l'\'Ecole Normale Sup\'erieure 
                          \vskip -12pt} \vskip 10pt
{\footnotesize
       \hoch{\phantom{2}} et \`a l'Universit\'e de Paris-Sud}} 

\pagebreak
\setcounter{page}{1}

\section{Introduction}

          BPS saturated $p$-brane solitons in supergravities play an
important role in non-perturbative string theory or M-theory, since
the supersymmetry which these solutions preserve may protect them from
non-perturbative quantum corrections in the theories.  These $p$-brane
solutions are described by harmonic functions on the space transverse
to the world-volume of the $p$-brane.  There are two different types
of dimensional reduction that may be used to relate $p$-brane
solutions of supergravity theories in different dimensions.  The
better known of these employs Killing symmetries of the $p$-brane
solutions in $D$ dimensions to effect a simultaneous reduction on the
world-volume and in the target space, yielding a $(p-1)$-brane in
$(D-1)$ dimensions.  This is the field-theoretic analogue of the
double dimensional reduction procedure for $p$-brane actions \cite{dhis},
which is also applicable to supergravity solutions, causing
``diagonal'' movements on the $D$ versus $p$ ``brane-scan.''  Aside
from a Weyl rescaling of the gravitational action, this diagonal
reduction procedure does not change the asymptotic fall-off behaviour
of a solution in the directions transverse to the world-volume.  In
other words, the harmonic functions that describe the $p$-branes in
different dimensions are preserved under diagonal dimensional
reduction.

        The second type of reduction has been referred to in the
literature as ``constructing periodic arrays'' \cite{mcenter,k,ghl}.
This procedure employs the zero-force property of extremal $p$-brane
solutions, which allows the existence of multi-centre solutions as
well as single-centre ones.  For example, let us consider an extremal
5-dimensional black hole solution, which is described by an harmonic
function $H$ on the 4-dimensional transverse space.  For a single
black hole located at the origin, we have $H=1 +\hat Q\, r^{-2}$.
Owing to the no-force condition between extremal black holes we can
construct multi-centre solutions, with, in particular, the black holes
arrayed periodically along a line:
\bea
H&=& 1 +\sum_{k=-\infty}^{\infty} \fft{\hat Q}{r^2 + (z+2\pi\,k\,
R)^2} \label{bsum}\\
&=& 1 + \fft{Q\, \sinh (r/R)}{2r\,
(\cosh (r/R) -\cos (z/R))}\ .\label{bsummed}
\eea
(The possibility of having a closed form for this summation was shown
in \cite{ghl}.)  Here, the 4-dimensional transverse space is described
by three coordinates $y^m$, with $r^2=y^m\, y^m$, together with the
coordinate $z$ along the line of black holes.  We have defined $Q$ in
terms of the 5-dimensional charge $\hat Q$ by $Q=\hat Q/R$.  If the
internal coordinate $z$ has exactly the same $2\pi\, R$ period as the
separation between the 5-dimensional black holes, then the solution
(\ref{bsum}) describes precisely a {\it single} black hole solution in
a space with a periodically identified (\ie compactified) coordinate
$z$.  The solution (\ref{bsum}) or (\ref{bsummed}) cannot be
dimensionally reduced on the $z$ coordinate in all the space since it
is $z$-dependent, contradicting the usual Kaluza-Klein requirement.
However, for large values of $r$, (or, equivalently, small values of
the period $2\pi R$), the $z$-dependence becomes negligible, and the
harmonic function $H$ tends to $1 + Q/(2r)$.  (Note that $Q$ therefore
has the interpretation of being the 4-dimensional charge.) In other
words, if we make a Fourier expansion of the solution (\ref{bsummed})
with respect to the compactified coordinate $z$, the non-zero modes
become insignificant in the regime $r>>R$, and only the zero-mode
survives.  Thus the solution becomes effectively a 4-dimensional black
hole.

    To make this precise, we may perform the Fourier transformation of 
(\ref{bsummed}) explicitly.  In an expansion that is convergent at
large $r$, it is not hard to see that this takes the form
\be
H = 1 +\fft{Q}{2r}\, \sum_m e^{-|m|\, r/R}\, e^{{\rm i} mz/R} =
1 +\fft{Q}{2r} +\fft{Q}{r}\, \sum_{m>0} e^{-m\, r/R}\,
\cos(\fft{m z}{R})\ , \label{exp}
\ee
which shows that the $z$-dependent terms tend to zero exponentially at
large $r$.  It is also of interest to consider a Fourier expansion
that is convergent in the regime where $r$ is small.  It is easy to see
that the appropriate expansion takes the form
\be
H=1 -\fft{Q}{r}\, \sum_{m>0} \sinh(\fft{mr}{R})\, \cos(\fft{m z}{R})
\ .\label{exps}
\ee 

     It is worth emphasising that in order to make the vertical
Kaluza-Klein dimensional reduction rigorously exact, it is necessary
to exploit the the no-force condition in the higher dimension in order
first to construct a {\it continuous} uniform distribution of charge
along the coordinate destined for Kaluza-Klein reduction
\cite{lpsvert}. Equivalently, this means choosing the relevant
harmonic function $H$ in the higher dimension to be independent of the
intended compactification coordinate.  Having then performed the
Kaluza-Klein reduction, one can always choose to carry out the inverse
procedure, and retrace the solution back to the higher dimension.
This inverse of Kaluza-Klein reduction, which has therefore acquired
the name ``Kaluza-Klein oxidation,'' leads back to the continuum
solution in the higher dimension, where the harmonic function is
independent of the compactification coordinate.  Thus under this
straightforward oxidation procedure, the vertical oxidation of a
$p$-brane in the lower dimension yields a continuous line of
$p$-branes in the higher dimension.  By contrast diagonal oxidation,
\ie the inverse of a diagonal Kaluza-Klein reduction, takes a single
lower-dimensional $p$-brane to a single higher-dimensional
$(p+1)$-brane.
        
        In general, let us consider a $D$-dimensional $p$-brane with
$\td d +2$ transverse dimensions.  The harmonic function for an
isotropic solution satisfies
\be
\square_{(\td d+2)} H = H'' + \fft{\td d +1}{r}\, H' =0\ ,\label{req}
\ee
where a prime denotes a derivative with respect to $r=\sqrt{y^m y^m}$.
The solution is therefore given by $H=1 + Q\, r^{-\td d}$.  If the
solution can be vertically oxidised to $D+1$ dimensions, the higher
dimensional harmonic function $\hat H$ will still retain this same
form, $\hat H = H$, and satisfy the same equation (\ref{req}), if we
simply retrace the step of Kaluza-Klein dimensional reduction.
However, we know that more general solutions for $\hat H$ are also
possible in $D+1$ dimensions, in which dependence on the extra
coordinate $z$ is also allowed.  In general, the function $\hat H$
must satisfy the $(\td d+3)$-dimensional harmonic condition
\be
\square_{(\td d+3)} \hat H = \hat H'' + \fft{\td d +1}{r}\, \hat H' +
\ddot {\hat H} =0\ ,\label{zeq}
\ee
where a dot denotes a derivative with respect to the compactification
coordinate $z$.  (We continue to assume that the function $\hat H$
depends on the lower-dimensional transverse coordinates $y^m$ only
through the isotropic distance $r=\sqrt{y^m\, y^m}$.)  There are
special solutions to the harmonic equation (\ref{zeq}) that include a
single $p$-brane solution, with $\hat H$ of the form $1 + \hat
Q\,(r^2+z^2)^{-(\td d +1)/2}$, and also a periodic array of
$p$-branes, of the form given in (\ref{bsum}), appropriately
generalised to a transverse space of the dimension $\td d+3$.  The
most general solution can be obtained by Fourier expanding the
dependence on the compactifying coordinate $z$, and writing $\hat
H=\sum_m f_m(r)\, e^{{\rm i} m z/R}$, where the functions $f_m(r)$ will
satisfy
\be
f_m'' + \fft{\td d +1}{r} f_m' - m^2 f_m =0\ .\label{beseq}
\ee
The solution to this equation is expressed in terms of the modified
Bessel functions $K$ and $I$,
\be
f_m= \fft{c_m}{r^{{\td d}/2}} \,  
K_{{\td d}/2}(\fft{|m|r}{R}) +  \fft{\tilde c_m}{r^{{\td d}/2}} \,  
I_{{\td d}/2}(\fft{|m| r}{R}) \ .\label{fexp}
\ee
For a series solution that is nonsingular at large $r$, we should
choose $\tilde c_m=0$ and retain only the $K$ Bessel functions.  Thus
the non-zero modes, {\it i.e.}\ $f_m(r)$ with $m\ne 0$, tend
exponentially to zero for large $r$.  The zero-mode solution $f_0$ is
precisely given by $H$, the lower-dimensional harmonic function.  The
$I$ Bessel functions are appropriate for an expansion valid at small
values of $r$.  The example of the 5-dimensional black hole that we
discussed previously fits into this general pattern, when we set $\td
d=1$.  In particular, since we have
\be
K_{1/2}(x) = \sqrt{\fft{\pi}{2x}}\, e^{-x}\ ,\qquad\qquad
I_{1/2}(x) = \sqrt{\fft2{\pi x}}\, \sinh x\ ,
\ee
we see that indeed the general form (\ref{fexp}) encompasses the
large-$r$ and small-$r$ expansions (\ref{exp}) and (\ref{exps}), with 
$c_m=Q\, \sqrt{m/(2\pi R)}$ or $\tilde c_m=-Q\,
\sqrt{m\pi/(8R)}$ respectively. 

      We have seen that all configurations for the higher-dimensional
solution can vertically reduced to lower-dimensional $p$-branes, as
long as the space is compactified, and that we focus attention only on
the regime where $r$ is large compared with the compactification
radius.  The construction of periodic arrays is not essential, as long
as the non-zero modes for any configuration die off at large values of
$r$.  In this description, the Kaluza-Klein reduction is approximate,
in that it can only be performed in the regime where $r$ is large.
For smaller values of $r$, the internal dimensions, and hence the
non-zero modes, become more and more important and the theory becomes
essentially higher-dimensional.  If one wishes to have a Kaluza-Klein
reduction that is exact everywhere, the solution should be constructed
using only the zero-mode, which can be achieved by taking a continuous
``stack'' of charges distributed uniformly around the internal
direction \cite{lpsvert}.

     To summarise, we may say that although the ``naive" vertical
oxidation of a single isotropic $p$-brane solution gives rise to a
continuum of $p$-branes in the higher dimension, a more natural
approach, from the point of view of solitons in string theory, is to
consider a periodic array of $p$-branes in the higher dimension.
Provided that one stays at distances large compared with the period,
the solution admits a lower dimensional interpretation that
approximates to the previously-discussed isotropic $p$-brane.

      In our discussion so far, we have considered the dimensional
reduction of a single $p$-brane.  In general there exist multi-charge
$p$-brane solutions, and the oxidation of such solutions will give
rise to intersections \cite{intsec} in the higher dimensions.
Multi-charge $p$-branes in $D\ge 2$ were classified in \cite{classp},
and their higher dimensional interpretations as intersections were
classified in \cite{bergsetal,classp}.  (See also a recent review
\cite{youm} on black holes and solitons.) To be precise, in higher
dimensions the standard intersecting solutions describe intersections
of $p$-branes whose charges are uniformly distributed on certain
compactifying coordinates, such that the solution is independent of
these coordinates.  However, just as in the case of the single-charge
$p$-branes we discussed previously, it is really more natural to look
for solutions that describe periodic arrays in the higher dimension,
rather than continuous distributions. In this paper we shall
investigate this question, by studying how the solutions can depend on
the compactifying coordinates (or relative transverse coordinates, if
one considers only the higher dimensional theory). This will provide
solutions for intersecting $p$-branes that go beyond those that are
normally considered.  As well as being more generic than the usual
intersecting solutions, they also provide a resolution of a previously
mysterious aspect of the extremal multi-charge $p$-brane solutions.
These are normally viewed as ``bound states'' of single-charge
$p$-branes \cite{boundstates}, with, however, zero binding energy.
What is not normally clear, therefore, is why the multi-charge
$p$-branes should have any particular physical significance in their
own right, since there appears to be nothing to prevent the individual
charge-centres from ``drifting apart,'' so that the configuration
separates into its single-charge constituents.  These multi-charge
$p$-branes are the dimensional reductions of the standard intersecting
solutions that involve continuous charge distributions in the higher
dimension.  If we instead consider the more generic intersections that
we find in this paper, then the interpretation as lower-dimensional
multi-charge $p$-branes is only approximate, in the regime where one
is far away from the charge centres.  The true solutions do not admit
any ``bound state at threshold'' interpretation except as a
large-distance approximation, and so the possibility of the charge
centres drifting apart no longer arises.

    Intersecting solutions of the more general kind we are considering
here have also been discussed in \cite{cvet1,cvet2,tseyt1,horo,callan,tseyt2}.

    In section 2 we study pair-wise intersections, and we find that
the form of the generalised solutions are similar to those obtained
directly from Kaluza-Klein oxidation, except that now the harmonic
functions generalise to become functions that ``interact'' with each
other.  We obtain general such solutions, discuss how they reduce to
lower dimensions, and we comment on their physical significance.  We
also present summarising rules for how to generalise the usual
intersecting solutions to the more general interacting ones.  In
section 3, we extend the discussion to multi-intersections, which can
be constructed by requiring that the conditions for all possible
pair-wise intersections be satisfied.  In section 4, we study the
supersymmetry of these generalised solutions.  We conclude the paper
in section 5.

           In the subsequent discussion we follow the notation of
\cite{lpsol} to denote the fields in $D$-dimensional maximal
supergravity, which are obtained by dimensional reduction from the 
eleven-dimensional fields $g_{\sst{MN}}$ and $A_{\sst{MNP}}$;
\bea
g_{\sst{MN}} &\longrightarrow & g_{\sst{MN}}\ ,\qquad \vec\phi\ ,\qquad
{\cal A}_1^{(i)}\ ,\qquad {\cal A}_0^{(ij)} \ ,\nn\\
A_3 &\longrightarrow & A_3\ ,\qquad A_2^{(i)}\ , \qquad A_1^{(ij)}\ ,
\qquad A_0^{(ijk)}\ ,\label{dfields}
\eea
where the superscript indices $i, j, k=1,\ldots, 11-D$ run over the
$11-D$ internal toroidally-compactified dimensions, starting from
$i=1$ for the step from $D=11$ to $D=10$, and the numerical subscripts denote
the degrees of the differential forms.  The bosonic Lagrangian in
$D$ dimensions is given by \cite{lpsol}
\bea
{\cal L} &=& eR -\ft12 e\, (\del\vec\phi)^2 -\ft1{48}e\, e^{\vec a\cdot
\vec\phi}\, F_\4^2 -\ft{1}{12} e\sum_i
e^{\vec a_i\cdot \vec\phi}\, (F_3^{(i)})^2
-\ft14 e\, \sum_{i<j} e^{\vec a_{ij}\cdot \vec\phi}\, (F_2^{(ij)})^2
\label{dgenlag}\\
&& -\ft14e\, \sum_i e^{\vec b_i\cdot \vec\phi}\, ({\cal F}_2^{(i)})^2
-\ft12 e\, \sum_{i<j<k} e^{\vec a_{ijk} \cdot\vec \phi}\,
(F_1^{(ijk)})^2 -\ft12e\, \sum_{i<j} e^{\vec b_{ij}\cdot \vec\phi}\,
({\cal F}_1^{(ij)})^2 + {\cal L}_{\sst{FFA}}\ ,\nn
\eea
where the ``dilaton vectors'' $\vec a$, $\vec a_i$, $\vec a_{ij}$,
$\vec a_{ijk}$, $\vec b_i$, $\vec b_{ij}$ are constants that
characterise the couplings of the dilatonic scalars $\vec \phi$ to the
various gauge fields.  The full details of these vectors, the
definitions of the field strengths, and the Wess-Zumino terms 
${\cal L}_{\sst{FFA}}$ can be found in \cite{lpsol}.

\section{Pair-wise intersections}

      In this section we discuss general solution of pair-wise
intersections, which are the dimensional oxidations of lower-dimensional
$p$-branes.  As in the case of the single-charge solution discussed in
the introduction, if we simply use the usual Kaluza-Klein ansatz for
the oxidation, the higher dimensional solution will necessarily be
independent of the internal compactifying space.  In the case where
the oxidation is vertical, it describes a configuration where the
higher-dimensional $p$-branes are uniformly distributed over the
internal space.  However in general the higher-dimensional solution
can also have non-vanishing non-zero modes, in which the functions $H$
acquire additional dependence on the internal coordinates, for which
the non-zero-modes tend to zero in the regime where the
lower-dimensional transverse radius is much larger than the
compactification scale.

        We shall first consider cases of double intersections arising
from the diagonal oxidation of one extended object together with the
vertical oxidation of the other extended object.  Let us consider for
example a two-charge black hole solution in D=4, with the charges
supported by the field strengths $F_2^{(ij)}$ and $F_2^{(k7)}$, where
the indices $i,j,k$ are all different, but chosen from the set
$\{1,2,\ldots,6\}$.  (The index value $i$ labels the internal
coordinate $z^i$ arising in the dimensional reduction from $12-i$ to
$11-i$ dimensions.)  The solution is given by
\bea
&&ds_4^2 = -(H_1 H_2)^{-1/2}\, dt^2 + (H_1 H_2)^{1/2}\, dy^m dy^m
\ ,\nn\\
&&\vec \phi = \ft12 \vec a_{ij}\, \log H_1 +\ft12 \vec a_{k7}\,
\log H_2\ ,\label{d4bh}\\
&&F_2^{(ij)}=dt\wdg dH_1^{-1}\ ,\qquad F_2^{(k7)} = dt \wdg
dH_2^{-1} \ ,\nn
\eea
where $H_1$ and $H_2$ are harmonic functions on the 3-dimensional
transverse space $y^m$, {\it i.e.} $\square H_1=0=\square H_2$, where
$\,\square = \del_m\del_m$.  Since $F_2^{(ij)}$ and $F_2^{(k7)}$ in $D=4$
arise from the dimensional reduction of $\hat F_2^{(ij)}$ and $\hat
F_3^{(k)}$ in $D=5$, it follows that the oxidation of the solution
(\ref{d4bh}) to $D=5$ gives a black hole (supported by $\hat
F_2^{(ij)}$) intersecting with a string (supported by $\hat F_3^{(k)}$),
{\it viz}.
\bea
&&ds_5^2 = \hat H_1^{-2/3}\, \hat H_2^{-1/3} \,(-dt^2 + 
           \hat H_1\, \hat H_2\, dy^m dy^m
           + \hat H_1\,  dz_7^2)\ ,\nn\\
&&\vec{\hat \phi} = \ft12 {\vec {\hat a}}_{ij}\, \log \hat H_1 +
                   \ft12 {\vec {\hat a}}_k\, \log \hat H_2
\ ,\label{d5sb}\\
&&\hat F_2^{(ij)} =dt\wdg d\hat H_1^{-1}\ ,\qquad
   \hat F_3^{(k)} =dt\wdg dz_7 \wdg d \hat H_2^{-1}
\ .\nn
\eea

     As in the example of the single-charge $p$-branes discussed in
the Introduction, here too we may entertain the idea that that the
functions $\hat H_1$ and $\hat H_2$ in $D=5$ need not simply be the
same harmonic functions $H_1$ and $H_2$ of the 4-dimensional 2-charge
solution, but might depend also on the compactification coordinate
$z_7$.  It is not so obvious in this 2-charge case, however, as to
what form this additional coordinate dependence might take.  Since the
string component is diagonally oxidised from $D=4$ to $D=5$, it is
natural to expect that $\hat H_2$ should continue to be an harmonic
function in $y^m$.  The black hole component, on the other hand, is
vertically oxidised to $D=5$, and naively, one might expect that $\hat
H_1$ could be an harmonic function on its entire transverse space
$\{z_7, y^m\}$ in $D=5$, as in the single-charge example.  This would
imply that there would be no force between the {\it isotropic} string
and black hole.  In fact it turns out that this naive guess does not
work.  We find indeed that more general solutions are in fact
possible, in which $\hat H_2$ continues to depend only on the
lower-dimensional transverse coordinates $y^m$, while the function
$\hat H_1$ is allowed to depend on $z_7$ as well as on $y^m$.  However
$\hat H_1$ and $\hat H_2$ do not decouple.  After calculations of some
complexity, we find that the 5-dimensional equations of motion are all
satisfied if $\hat H_1$ and $\hat H_2$ satisfy the equations
\be
\square\, \hat H_2 = 0 \ ,\qquad
\square\, \hat H_1 + \hat H_2\, \ddot {\hat H_1} =\ 0,\label{intseceom1}
\ee
where $\square$ again denotes the Laplacian in the $y^m$ coordinates,
and a dot denotes a derivative with respect to $z_7$.  Note that $\hat
H_2$ is still a harmonic function on the 3-dimensional $y^m$ space,
just like $H_2$ in the lower dimension, but $\hat H_1$ is no longer a
harmonic function in general.  (This implies that in the presence of a
string, there can be no static isotropic black hole.)  The metric,
dilatons and field strengths are still given by (\ref{d5sb}), in terms
of the new functions $\hat H_1$ and $\hat H_2$ that satisfy
(\ref{intseceom1}).

    The equations in (\ref{intseceom1}) admit two special solutions, 
in which $\hat H_1$ and $\hat H_2$ decouple, namely
\medskip

1)\  {\it $\hat H_1=H_1(y)$ and $\hat H_2=H_2(y)$ are both harmonic on
the overall transverse space $y^m$.}
\medskip

2)\  {\it $\hat H_1=\hat H_1(z_7)$ and $\hat H_2=\hat H_2(y)$ are
harmonic on $z_7$ and $y^m$ respectively.}
\bigskip

\noindent Note that the first special solution is independent of
$z_7$, and the solution can be simply reduced to the $D=4$ solution
(\ref{d4bh}).  Conversely, it is the solution in $D=5$ that is
obtained simply by the standard oxidation of the 2-charge black hole
in $D=4$.  The second special solution cannot be dimensionally reduced
on the $z_7$ coordinate, on account of the $z_7$ dependence of $\hat
H_1$. This type of special solution was also found in
\cite{bergsetal}.

         We are interested in finding more general classes of
solutions to (\ref{intseceom1}), where, however, the non-zero modes
become negligible in the regime where the radius $r=\sqrt{y^m\, y^m}$
is large compared the radius of the compactifying coordinate.  To
construct such solutions, we consider the situation where $\hat H_2$
is still an isotropic harmonic function, {\it i.e.}\ $\hat H_2 = 1 +
\hat Q_2/r$. In other words, we study the black hole configuration in
the presence of a single string, which, without loss of generality,
can be located at $y^m=0$.\footnote{Of course here, and for the similar
equations we shall encounter later in the paper, more general
solutions of the equations (\ref{intseceom1}) are also possible, where
we do not assume only isotropic dependence on the $y^m$ coordinates.}  
We shall also take $\hat H_1$ to depend on
$y^m$ isotropically through the coordinate $r$, but we allow it also
to depend on $z_7$.  (Note that since $\hat H_2$ is already assumed to
be isotropic, this implies that only the $z_7$-independent part ({\it
i.e.}\ the zero-mode) of $H_1$ could in any case naturally have any
non-isotropic dependence on $y^m$.)  It then follows from
(\ref{intseceom1}) that $\hat H_1$ satisfies the equation
\be
\square\, \hat H_1 + H_2\, \ddot {\hat H_1}=
H_1'' + \fft{2}{r} H_1' + (1 + \fft{\hat Q_2}{r})\, \ddot {\hat H_1}
=0\ .\label{int1}
\ee
It is worth emphasising again that for non-vanishing charge $\hat
Q_2$, the function $\hat H_1$ is {\it not} an harmonic function on the
4-dimensional transverse space $\{z_7, y^m\}$.  However, in the
asymptotic regime where $r>>\hat Q_2$, the harmonic function $\hat
H_2$ tends to unity, and then $\hat H_1$ becomes an harmonic function
in its traverse space. Thus an observer at $r>>\hat Q_2$ would
conclude that $\hat H_1$ is an harmonic function on its 4-dimensional
transverse space $\{z_7, y^m\}$, which could be isotropic as if there
existed an extremal 5-dimensional black hole.  One can therefore
construct a periodic array of such solutions (\ref{bsum}) aligned
along the $z_7$ axis, for which the $z_7$-dependence will become
negligible at large $r$.

    An observer in the lower dimension who observed the black-hole
configuration from large distances $r$ would conclude that there was a
no-force condition between the two charge species, by virtue of the
independent harmonic nature of the two functions $H_1$ and $H_2$.  At
large $r$, he would be unable to distinguish between the situation
where the higher-dimensional functions $\hat H_1$ and $\hat H_2$ were
also exactly harmonic, and the situation where they have the more
general interacting form that we have been considering here.  If they
do interact, with $\hat H_1$ being a non-harmonic solution of the more
general equation (\ref{int1}), then the no-force condition in the
lower dimension, with its consequence that the two charge centres
could separate freely, would be a pure artefact of the large-distance
approximation, and a closer inspection of the small-$r$ region near to
the black hole would reveal a more complex higher-dimensional
structure.

     We shall now show that indeed such a more general configuration
can also be dimensionally reduced to $D=4$, and give rise to
(\ref{d4bh}), provided that we are in the regime $r>>R_7$, where $R_7$
is the radius of the compactification coordinate $z_7$.  To do this,
we make a Fourier expansion of $\hat H_1$ on $z_7$, writing $\hat H_1
= \sum_m f_m(r) e^{{\rm i} m z_7/R_7}$.  The Fourier modes $f_m(r)$
then satisfy
\be
f_m'' + \fft{2}{r} f_m' - \fft{m^2}{R^2} (1 + \fft{\hat Q_2}{r}) f=0
\ .
\ee
The general solution to this equation is given by
\be
f_m =c_m\, U(1+\fft{|m|\hat Q}{2R_7}, 2, \fft{2|m|r}{R_7})\,
e^{-|m|r/R_7} +
    \tilde c_m\, M(1+\fft{|m|\hat Q}{2R_7}, 2, \fft{2|m|r}{R_7})
\, e^{-|m|r/R_7}\ ,\label{kummer}
\ee
where $U(a,b,x)$ and $M(a,b,x)$ denote Kummer's confluent
hypergeometric functions.  We are interested for now in expansions
that are non-singular at large $r$, implying that we should take
$\tilde c_m=0$ and retain only the $U(a,b,x)$ functions.  The non-zero
modes ($f_m$ with $m\ne0$) in fact tend to zero for large $r/R$, and
the zero-mode solution $f_0$ is precisely given by $H_1$, the
lower-dimensional harmonic function.  Thus we see that in the regime
$r>>R_7$, the solution of the non-linear equation of motion can also
be dimensionally reduced, since its dependence on $z_7$ becomes
negligible, and it limits to the previous $z_7$-independent form.
Note that this description in terms of a Fourier expansion where the
non-zero modes tend to zero at large $r$ is precisely analogous to the
description we gave for single $p$-branes in the Introduction.
Indeed, if we set $\hat H_2$ to be unity in the above discussion, the
series (\ref{kummer}) reduces precisely to the form of the expansion
(\ref{fexp}), for the particular case $\td d=1$.

      As was observed in \cite{boundstates}, the two-charge black hole
solution (\ref{d4bh}) can be viewed as a bound state of two basic
single-charge black-hole building blocks, however with zero binding
energy.  This latter feature implies that the two individual building
blocks can move freely with respect to one another.  This picture
makes the existence of the two-charge black hole (\ref{d4bh}) in $D=4$
appear quite accidental, since the two charge centres can apparently
freely drift apart.  However, as we have shown, if we interpret such a
solution from a higher-dimensional point of view, the non-interacting
(harmonic) functions $H_1$ and $H_2$ in $D=4$ become functions $\hat
H_1$ and $\hat H_2$ that in general do interact with each other, and
hence the relative locations of the two charges are no longer
arbitrary; they are located at the same point in the overall
transverse space of the $y^m$ coordinates, indicating that the
individual components in the four-dimensional 2-charge solution are in
some sense bound together and cannot drift apart.  In fact the
interpretation of the configuration as a bound state of two basic
objects is no longer appropriate for these more generic interacting
solutions.  (However, it is worth observing that there are no {\it
self-interactions} of the individual $\hat H$ functions, implying that
there is no force between $p$-branes carried by the same field
strength.) Of course there do also exist configurations in $D=5$ where
the two intersecting objects are genuinely independent.  This happens
when $\hat H_1$ and $\hat H_2$ are both taken to have one or other of
the two special forms discussed previously.  The first of these cases,
where the functions have the coordinate dependences $\hat H_1(y)$ and
$\hat H_2(y)$, implies that there is no force between a string and a
black-hole configuration whose black hole charge is distributed
uniformly along the string world-line.  The second case, with $\hat
H_1(z_7)$ and $\hat H_2(y)$, implies that there is no force between a
string and a black-hole configuration whose charge is uniformly
distributed over the entire space transverse to the string.

        The above discussion applies to other pair-wise intersections
as well.  Let us look again at a similar example where upon oxidation,
one extended object undergoes vertical oxidation, whilst the other
undergoes diagonal oxidation.  We shall consider an NS-NS dyonic string
in $D=6$, supported by the NS-NS 3-form field strength $F_3^{(1)}$.  When
it oxidises to $D=10$, it becomes the intersection of an NS-NS string and
a 5-brane, with the solution in $D=10$ given by
\bea
ds_{10}^2 &=& \hat H_e^{-3/4} \hat H_m^{-1/4}\, (-dt^2 + dx_1^2 + H_e H_m\,
dy^m dy^m + H_e\, (dz_2^2 + \ldots + dz_5^2) )\ ,\nn\\
\phi &=& \ft12 \log (\hat H_e/\hat H_m)\ ,\label{d10s5}\\
F_3^{(1)} &=& dt\wdg dx_1 \wdg dH_e^{-1} +
              e^{-\phi}\, *(dt \wdg dx_1 \wdg d^4z \wdg dH_m^{-1})
\ .\nn
\eea
In the directly oxidised solution, the functions $H_e$ and $H_m$ depend
only on the four transverse coordinates $y^m$ of the
original 6-dimensional solution.  Again, however, we may consider more
general types of solution in $D=10$ where appropriate dependence on the
internal compactification coordinates $\vec z$ is also allowed. 
Specifically, we find that such more general solutions exist in $D=10$,
in which the metric, dilaton and field strength are still given in terms
of $\hat H_e$ and $\hat H_m$ by (\ref{d10s5}), but now $\hat H_e$ is
allowed to depend on the internal coordinates $\vec z$ as well as
on $y^m$. We find that the ten-dimensional equations of motion are
all satisfied provided that the functions $\hat H_e(y,z)$ and $\hat
H_m(y)$ satisfy
\be
\square_{\vec y} \,\hat H_m =0\ ,\qquad
\square_{\vec y}\, \hat H_e + \hat H_m \, \square_{\vec z}\,\hat  H_e=0
\ ,\label{s5inteom}
\ee
where $\square_{\vec y}$ and $\square_{\vec z}$ denote the flat-space
Laplacians in the four $y^m$ and the four $z^i$ coordinates
respectively.  

      Again these equations admit two special solutions analogous to
those discussed in the previous example, namely a solution where $\hat
H_e$ is also independent of $\vec z$, and so both functions are simply
harmonic in $\vec y$, and a solution where $\vec H_e$ depends {\it
only} on $\vec z$, and the functions $\hat H_e$ and $\hat H_m$ are
harmonic in $\vec z$ and in $\vec y$ respectively.  The former
corresponds to the simple oxidation of the $D=6$ dyonic string, while
the latter corresponds to a special kind of solution described in
\cite{bergsetal}.  We are interested instead in the general solutions
that can be obtained by performing a Fourier expansion of $\hat H_e$
on the compactification coordinates, $\hat H_e(\vec y,\vec
z)=\sum_{\vec n} f_{\vec n}(\vec y)\, e^{{\rm i} n_i\, z^i/R_i}$,
where $R_i$ denotes the radius of the compactification circle for the
coordinate $z^i$, and $\vec n$ denotes the set of integers
$\{n_i\}$. The Fourier modes in this case satisfy
\be
f_{\vec n}'' + \fft{3}{r}\, f_{\vec n}' - \lambda_{\vec n}^2\, 
(1+\fft{Q_m}{r^2})\, f_{\vec n} =0\ ,
\ee
where we consider functions $\hat H_e(r,\vec z)$ and $\hat H_m(r)$ that
depend isotropically on $\vec y$, we take $H_m=1+Q_m/r^2$, and
\be
\lambda_{\vec n} =\sqrt{\sum_i n_i^2/R_i^2}\ .
\ee 
The solutions for the Fourier modes are then given in terms of
modified Bessel functions by
\be
f_{\vec n} = \fft{c_{\vec n}}{r}\, K_\a(\lambda_{\vec n}\, r) +
   \fft{\tilde c_{\vec n}}{r}\, I_{\a}(\lambda_{\vec n}\, r)\ ,
\ee
where $\a=\sqrt{1+\lambda_{\vec n}^2\, Q_m}$.  In order to have
expansions that are nonsingular at large $r$, we must take $\tilde
c_{\vec n}=0$, so that the solutions are given in terms of the $K$
Bessel functions only. It is easily verifiable that in the regime
where $r >>R_i$ for all $i$ the non-zero modes become negligible, and
the solution effectively becomes the usual oxidation of the
six-dimensional dyonic string.

           In the above two examples, we considered cases where the
internal space is the world-volume for one of the two extended objects. 
For a generic multi-charge $p$-brane, each individual object will
undergo a vertical oxidation in general.  For example, if we oxidise
the above two examples further to $D=11$, both extended objects in
the intersection will undergo further vertical oxidations.  In
particular, if we oxidise the intersection of the 10-dimensional string
and 5-brane further to $D=11$, the 5-brane will now be vertically
lifted, given to an intersection of membrane and 5-brane in $D=11$:
\bea
ds_{11}^2 &=& \wtd H_e^{-2/3} \wtd H_m^{-1/3}\, \Big( -dt^2 + dx_1^2 +
               \wtd H_e \wtd H_m\, dy^m dy^m \nn\\
          && \qquad\qquad
+ \wtd H_m\, dz_1^2 + \wtd H_e\, (dz_2^2 + \cdots + dz_5^2)
\Big) \ ,\label{d11m5}\\
F_4 &=& dt\wdg dx_1 \wdg dz_1 \wdg d\wtd H_e^{-1} +
      *(dt\wdg dx_1 \wdg dz_2\wdg\cdots\wdg dz_5\wdg 
         d\wtd H_m^{-1})\ ,\nn
\eea
In this case, the associated harmonic functions $\hat H_e
\rightarrow \wtd H_e$ and $\hat H_m \rightarrow\wtd H_m$ will satisfy 
either
\be
\square_{\vec y} \,\wtd H_m =0\ ,\qquad
\square_{\vec y}\, \wtd H_e + \wtd H_m \, \square_{\vec z}\,\wtd  H_e=0
\ ,\label{s6inteom1}
\ee
or
\be
\square_{\vec y}\, \wtd H_e =0\ , \qquad
\square_{\vec y}\, \wtd H_m + \wtd H_e\, \square_{z^1}\, \wtd H_m =0\ ,
\ee

         If we oxidise the first example, of the 4-dimensional
2-charge black hole (\ref{d4bh}) to $D=11$, it becomes the
intersection of two membranes.  It is a special case of more general
solutions that we can construct by starting from a 2-charge black-hole
in $D=7$, supported by field strengths coming from the 4-form in
$D=11$.  Without loss of generality, we may consider the case using
field strengths $F_2^{(12)}$ and $F_2^{(34)}$, with associated
harmonic functions $H_1(\vec y)$ and $H_2(\vec y)$.  We shall use the
notation that indices $a_1,b_1,\ldots$ run over 1 and 2, while indices
$a_2,b_2\ldots$ run over 3 and 4.  Having oxidised the
seven-dimensional solution to $D=11$, we then generalise the harmonic
functions to $\wtd H_1=\wtd H_1(\vec y, z^{a_2})$ and $\wtd H_2 = \wtd
H_2(\vec y, z^{a_1})$.\footnote{The general principle here, and in
all other cases, governing the ``extra'' coordinate dependences that
we consider for the $\wtd H$ functions is that we allow the function
associated with each lower-dimensional single-charge $p$-brane
component to depend on all its own transverse coordinates in the higher
dimension, \ie on all the coordinates associated with vertical
oxidation steps for that particular $p$-brane component (together with
the original transverse coordinates in the lower dimension).
Equivalently, we may state the conditions directly in the higher
dimension, without reference to oxidation, as the assumption that
each function in the intersection depends, {\it a priori}, on all its
transverse coordinates.  Equations of motion, as we shall see
below, may impose further restrictions.}
We find that the eleven-dimensional configuration
\bea
ds_{11}^2 &=& (\wtd H_1 \wtd H_2)^{-2/3}\, (-dt^2 + 
\wtd H_1 \wtd H_2\, dy^m
dy^m + \wtd H_1\,(dz_3^2 + dz_4^2) + \wtd H_2\, 
(dz_1^2 + dz_2^2))\ ,\nn\\
F_4&=& dt\wdg dz_1\wdg dz_2 \wdg d\wtd H_1^{-1} +
       dt\wdg dz_3\wdg dz_4\wdg d\wtd H_2^{-1}\ .
\eea
solves the supergravity equations of motion provided that the
functions $\wtd H_1$ and $\wtd H_2$ satisfy
\bea
\square_{\vec y}\, \wtd H_1 + \wtd H_2\, \del_{a_2}^2\, \wtd H_1 &=&0
\ ,\nn\\
\square_{\vec y}\, \wtd H_2 + \wtd H_1\,\del_{a_1}^2\, \wtd H_2 &=&0\ ,
\label{d11m2}
\eea
and also the constraint
\be
\epsilon_{a_1b_1}\, \del_{b_1}\wtd H_2\, \del_{a_2} \wtd H_1
-\epsilon_{a_2b_2}\, \del_{a_1}\wtd H_2\, \del_{b_2} \wtd H_1 =0\ .
\label{cons111}
\ee
This latter condition is equivalent to the constraint
\be
\del_{a_1} \wtd H_2\, \del_{a_2} \wtd H_1=0\ ,
\ee
which follows from (\ref{cons111}) upon using the fact that $\wtd H_1$
and $\wtd H_2$ are real.  Thus we have either $\wtd H_1=\wtd H_1(\vec
y)$ and $\wtd H_2=\wtd H_2(\vec y,z_1,z_2)$ or else  $\wtd H_1=\wtd H_1(\vec
y, z_3,z_4)$ and $\wtd H_2=\wtd H_2(\vec y)$.  Correspondingly, the
remaining equations (\ref{d11m2}) become
\be
\square_{\vec y}\, \wtd H_1 = 0\ ,\qquad  \square_{\vec
y}\, \wtd H_2 + \wtd H_1\, \del_{a_1}^2\, \wtd H_2 =0\ ,
\ee
or
\be
\square_{\vec y}\, \wtd H_2 =0\, \qquad  \square_{\vec y}\, \wtd H_1 +
\wtd H_2\, \del_{a_2}^2\, \wtd H_1 =0\ .
\ee

       To summarise, the spacetime of any double intersections of
$p$-branes is split into overlapping world volume coordinates $x^\mu$,
relative transverse coordinates $\vec z_1$ and $\vec z_2$, and overall
transverse space coordinates $\vec y$.  Assuming the
two intersecting objects, described by functions $H_1$ and $H_2$, have
world volumes $\{x^\mu, \vec z_1\}$ and $\{x^\mu, \vec z_2\}$
respectively, then these two functions satisfy
\be
\square_{\vec y}\, H_1 = 0\ ,\qquad
\square_{\vec y}\, H_2 + H_1\, \square_{\vec z_1}\, H_2 =0
\ ,\label{geneom1}
\ee
or
\be
\square_{\vec y}\, H_2 =0\ ,\qquad 
\square_{\vec y}\, H_1 + H_2\, \square_{\vec z_2}\, H_1 =0
\ .\label{geneom2}
\ee
Thus we see that in general the functions $H_1$ and $H_2$ interact
with each other.  Two special cases arise, where either the two
functions both depend only on the overall transverse space $\vec y$
(this is the usual kind of intersecting solution), or else one function
depends on the relative transverse space while the other depends on
the overall transverse space.  This latter special case was found in
\cite{bergsetal}.  In both cases, the $H_1$ and $H_2$ becomes
independent harmonic functions in the relevant coordinates, implying a
no-force condition.  The first case describes $p$-branes with their
charges uniformly distributed along each other's relative
world-volumes, while the second case describes one extended object
with charges uniformly distributed along the overall transverse space
of the other.  Only in these two circumstances do the forces between
the two objects cancel out.  We can obtain more general solution to
(\ref{geneom1}) or (\ref{geneom2}) by taking one of the functions to
be harmonic in $\vec y$, and the other to depend on $\vec y$
and the appropriate $\vec z$ coordinates.

\section{Multiple intersections}

           Multi-intersections in M-theory or string theories can be
obtained from the dimensional oxidation of multi-charge $p$-brane
solutions in lower dimensions.  Having constructed the pair-wise
intersections, the discussion for all multi-intersections follows
straightforwardly.  The only criterion for multiple intersections is
that they should reduces to the known pair-wise solutions for all
possible cases where all but two of the charges are set to zero.

           The example we shall examine is the intersections arising
from RN (Reissner-Nordstr{\o}m) black holes in $D=5$, which can be
viewed as 3-charge black holes in the $D=5$ maximal supergravity
theory. There are a total of 45 possible field configurations that
gives rise to such RN black holes in $D=5$, which form a
45-dimensional representation of the Weyl group of $E_6$
\cite{lpsweyl}.  We shall consider an example that uses the field
strengths $\{F_2^{(12)}, F_2^{(34)}, F_2^{(56)}\}$, for which the
5-dimensional black hole solution is given by
\bea
ds_5^2 &=& -(H_1H_2H_3)^{-2/3}\, dt^2 + (H_1 H_2 H_3)^{1/3}\, dy^m dy^m
\ ,\nn\\
\vec\phi &=& \ft12 \vec a_{12}\, \log H_1 + \ft12 \vec a_{34}
\log H_2 + \ft12 \vec a_{56}\, \log H_3 \ ,\label{d5bh}\\
F_2^{(12)} &=& dt\wdg dH_1^{-1}\ ,\quad
F_2^{(34)} = dt\wdg dH_2^{-1}\ ,\quad
F_2^{(56)} = dt\wdg dH_3^{-1}\ ,\nn
\eea
where $H_i$ are harmonic functions on 4-dimensional transverse space
$y^m$.  Let us first oxidise the solution to $D=6$.  The field
strengths $\{F_2^{(12)},F_2^{(34)}, F_2^{(56)}\}$ in $D=5$ are
dimensionally reduced from the those $\{\hat F_2^{(12)},\hat
F_2^{(34)}, \hat F_3^{(5)}\}$ in $D=6$. The solution in $D=6$
describes an intersection of 2-charge black hole with a string:
\bea
ds_6^2 &=& (\hat H_1 \hat H_2)^{-3/4} H_3^{-1/2}\, (-dt^2 +
\hat H_1 \hat H_2 \hat H_3\,  dy^m dy^m + \hat H_1 \hat H_2\, dz_6^2)
\ ,\nn\\
\vec {\hat\phi}&=& \ft12 {\vec {\hat a}}_{12}\, \log \hat H_1 +
               \ft12 {\vec {\hat a}}_{34}\, \log \hat H_2 +
               \ft12 {\vec {\hat a}}_5\, \log \hat H_3\ ,\label{d6bh2s}\\
\hat F_2^{(12)} &=& dt\wdg d\hat H_1^{-1}\ ,\quad
\hat F_2^{(34)} = dt\wdg d\hat H_2^{-1}\ ,\quad
\hat F_3^{(5)} = dt\wdg d\hat H_3^{-1}\ .\nn
\eea
It is a solution provided that 
\be
\square_{\vec y}\, \hat H_3 = 0\ ,\quad
\square_{\vec y}\, \hat H_1 + H_3\, \ddot {\hat H_1}=0\ ,\quad
\square_{\vec y}\, \hat H_2 + H_3\, \ddot {\hat H_2}=0
\ .\label{d6bn2seom1}
\ee
Note that in $D=6$, the solution should really be viewed as a double
intersection of a string and a two-charge black hole.  We are
interested in particular a class of solution where $H_3$ is taken to
be isotropic, {\it i.e.}\ $H_3 = 1 + \hat Q_3/r^2$.  Then $\hat H_i$
with $i=1,2$ 
can be Fourier expanded as $H_i =\sum_m f^i_m(r) e^{i m z_6/R_6}$, where
\bea
(f^i_m)'' + \fft3r (f^i_m)' - \fft{m^2}{R_6^2} (1 +\fft{\hat
Q_3}{r^2}) f_m^i =0 \ .
\eea
This has the solution
\be
f^i_m = \fft{c^i_m}{r}\, K_\a(\fft{m r}{R_6}) +
\fft{\tilde c^i_m}{r}\, I_\a(\fft{m r}{R_6}) \ ,
\ee
where $\a=\sqrt{1+ m^2\, \hat Q/R_6^2}$, and as usual we choose
$\tilde c^i_m=0$ and retain only the $K$ Bessel functions if we want
expansions that are non-singular as $r$ tends to infinity.

      It is important to note that the RN black hole has non-vanishing
entropy $S\sim \sqrt{Q_1 Q_2 Q_3}$ despite the fact that it is
extremal.  This entropy can be understood from a six-dimensional
viewpoint, where, for a suitable choice of the three supporting field
strengths in $D=5$, the solution becomes a ``boosted'' dyonic D-string
\cite{vs}.  (This occurs when one of the three 2-form field strengths
in $D=5$ is chosen to be the Kaluza-Klein 2-form coming from the
six-dimensional metric.  Such a choice of three field strengths is
related to any other valid choice (such as the one we have taken) by
U-duality.)  From the lower-dimensional point of view, owing to the
no-force condition between the charges, the existence of such a
solution appears ``accidental'', and the entropy would vanish if the
three building-block components of the RN black hole drifted apart.
The solution we have obtained here indicates that from the
higher-dimensional point of view, these three building blocks are
intrinsically bound together within the overall transverse space, {\it
i.e.}\ the 4-dimensional space in $D=5$.

        The solution can be further oxidised back to $D=11$, where it
becomes the intersection of three membranes:
\bea
ds_{11}^2 &=& (\wtd H_1 \wtd H_2 \wtd H_3)^{-2/3}\, \Big (-dt^2 +
                \wtd H_1 \wtd H_2 \wtd H_3\, dy^m dy^m\nn\\
          && + \wtd H_2 \wtd H_3\, (dz_1^2 + dz_2^2)
             + \wtd H_1 \wtd H_3\, (dz_3^2 + dz_4^2)
             + \wtd H_1 \wtd H_2\, (dz_5^2 + dz_6^2)\Big)\ ,\label{d11m3}\\
F_4 &=& dt\wdg d^2z_{12} \wdg d\wtd H_1^{-1} +
      dt\wdg d^2z_{34} \wdg d\wtd H_2^{-1} +
      dt\wdg d^2z_{56} \wdg d\wtd H_3^{-1}\ .\nn
\eea
Under the straightforward oxidation, the three $\wtd H$ functions
would simply be the same as the functions $\hat H$ in $D=6$.  However,
we can consider yet more general coordinate dependence now in $D=11$.
Specifically, we may consider the following possibility:
\be
\wtd H_1 = \wtd H_1(\vec y,z^{a_2},z^{a_3})\ ,\qquad
\wtd H_1 = \wtd H_1(\vec y,z^{a_1},z^{a_3})\ ,\qquad
\wtd H_1 = \wtd H_1(\vec y,z^{a_1},z^{a_2})\ ,\label{3zc}
\ee
where indices $a_1,b_1,\ldots$ run over 1 and 2, indices
$a_2,b_2,\ldots$ run over 3 and 4, and $a_3,b_3,\ldots$ run over 5 and
6.  We find that (\ref{d11m3}) is then a solution of $D=11$
supergravity, provided that the functions $\wtd H$ satisfy the
following conditions:
\bea
\square_{\vec y}\,\wtd H_1 
+ \wtd H_2\, \del_{a_2}^2 \wtd H_1 
+ \wtd H_3\, \del_{a_3}^2 \wtd H_1 &=& 0\ ,\nn\\
\square_{\vec y}\,\wtd H_2 
+ \wtd H_3 \,\del_{a_3}^2 \wtd H_2 
+ \wtd H_1 \,\del_{a_1}^2 \wtd H_2 &=& 0\ ,\label{harm} \\
\square_{\vec y}\,\wtd H_3 
+ \wtd H_1\, \del_{a_1}^2 \wtd H_3 
+ \wtd H_2\, \del_{a_2}^2 \wtd H_3 &=& 0\ ,\nn
\eea
and in addition the constraints
\bea
\epsilon_{a_1b_1}\, \del_{b_1}\wtd H_2\, \del_{a_2} \wtd H_1
-\epsilon_{a_2b_2}\, \del_{a_1}\wtd H_2\, \del_{b_2} \wtd H_1 &=&0\
 ,\nn\\
\epsilon_{a_2b_2}\, \del_{b_2}\wtd H_3\, \del_{a_3} \wtd H_2
-\epsilon_{a_3b_3}\, \del_{a_2}\wtd H_3\, \del_{b_3} \wtd H_2 &=&0\
 ,\label{ct0}\\
\epsilon_{a_3b_3}\, \del_{b_3}\wtd H_1\, \del_{a_1} \wtd H_3
-\epsilon_{a_1b_1}\, \del_{a_3}\wtd H_1\,\del_{b_1}\wtd H_3 &=&0\ .
\nn
\eea
(These equations (\ref{harm}) and (\ref{ct0}) can all be derived from
the field equation for $F_4$ in $D=11$; having imposed them, the
Einstein equation is also satisfied.)  The constraints (\ref{ct0}) can
be re-expressed in the simpler form
\bea
\del_{a_1} \wtd H_2 \, \del_{a_2} \wtd H_1 &=& 0\ ,\nn\\
\del_{a_2} \wtd H_3 \, \del_{a_3} \wtd H_2 &=& 0\ ,\label{ct}\\
\del_{a_3} \wtd H_1 \,  \del_{a_1} \wtd H_3 &=& 0\ ,\nn
\eea
as a consequence of the fact that the $\wtd H$ functions are real. 
Note that the equations (\ref{harm}) and the constraints (\ref{ct0})
for three intersections are just the natural extensions of
(\ref{d11m2}) and (\ref{cons111}) for double intersections (\ref{d11m2}).

     The conditions (\ref{ct}) impose restrictions, similar to those
we have already encountered in our other examples, on the possible
dependences of the $\wtd H$ functions on the internal compactification
coordinates.  There are clearly a variety of ways in which the
conditions (\ref{ct}) can be satisfied; provided this is done, then we
will have 3-intersection solutions as long as the functions $\wtd H$
satisfy now-familiar types of interacting equations, namely
(\ref{harm}).  One example of an allowed choice of coordinate
dependences is to take
\be
\wtd H_1=\wtd H_1(\vec y,z_3,z_4)\ ,\qquad \wtd H_2=\wtd
H_2(\vec y,z_5,z_6)\ ,
\qquad \wtd H_3=\wtd H_3(\vec y,z_1,z_2)\ ,\label{choice1}
\ee 
leading to quite complicated interactions between the functions, as
can be seen from (\ref{harm}).  Another inequivalent allowed choice for
the coordinate dependences is 
\be
\wtd H_1=\wtd H_1(\vec y)\ ,\qquad 
\wtd H_2=\wtd H_2(\vec y,z_1,z_2) \ ,
\qquad \wtd H_3=\wtd H_3(\vec y,z_1,z_2,z_3,z_4)\ .\label{choice2}
\ee 
In this case, $\wtd H_1$ will satisfy the original harmonicity
condition in $\vec y$, while both $\wtd H_2$ and $\wtd H_3$ will
satisfy more general equations in terms of Bessel
functions.\footnote{While we were completing the writing of this
paper, a preprint appeared that discusses some related ideas for
generalised forms of intersecting $p$-brane solutions
\cite{ivanov}. However, the configuration of three intersecting
membranes in $D=11$ proposed in \cite{ivanov}, with $\wtd H_1=
H_{11}(z^{a2}) H_{12}(z^{a_3}) H_{13}(\vec y)$ and cyclic
permutations, violates the constraints (\ref{ct0}) or (\ref{ct}).
Thus such configurations evidently conflict with the equations of motion.}

       Thus we note that the conditions for having
triple-intersections follow from the conditions for all possible
sub-combinations of pair-wise intersections.  These rules generalise
to higher numbers of intersections also.

\section{Supersymmetry}

     The fractions of supersymmetry preserved by the various solutions
we have been discussing here can be determined by standard means.  In
general, the simplest way to do this is first to oxidise the solution
to $D=11$, where the supersymmetry transformation rules are most
easily analysed.  Exploiting the fact that supersymmetry is preserved
under Kaluza-Klein dimensional reduction, the supersymmetry fraction
found in $D=11$ will be the same as that in the lower dimension.

     Our conclusions about the preserved fractions of supersymmetry
are easily summarised: for the generalised solutions we are
considering in this paper, where certain of the functions $H_i$ in an
intersecting solution acquire additional dependence on certain of the
internal coordinates, the fractions of preserved supersymmetry are
always the same as they were in the corresponding ``standard''
solutions.  Thus in particular, any generalised solution involving two
functions $H_i$ preserves $\ft14$ of the supersymmetry, and any
solution involving three functions preserves $\ft18$ of the
supersymmetry.

     We shall not present the complete details of the supersymmetry
calculations for all the solutions that we have obtained in this
paper; instead, we shall simply give some of the key steps.  One
necessary ingredient is the covariant derivative acting on spinors,
\ie $D_\mu\psi = \del_\mu\psi +\ft14 \omega_\mu^{ab}\,
\Gamma_{ab}\psi$.  It is useful to find the form of $D_\mu$ for a
rather general class of metrics, which we may represent in the form
\be
ds^2 = -e^{2u}\, dt^2 + e^{2v}\, dy^i\, dy^i + \sum_\A e^{2w^{\A}}\,
dz^\A\, dz^\A\ ,
\ee
where the functions $u$, $v$ and $w^\A$ depend on $y^i$ and $z^\B$. 
After calculating the spin connection in the obvious orthonormal basis 
$e^0=e^u\, dt$, $e^i=e^v\, dy^i$, $e^\A=e^{w^\A}\, dz^\A$, we find that
the tangent-space components of the spinor covariant derivative, $D_a =
e^\mu_a\, D_\mu$, are given by
\bea
D_0&=& e^{-u}\, \fft{\del}{\del t} +\ft12 e^{-v}\, u_i\, \Gamma_{0i}
  +\ft12 e^{-w^\A}\, u_\A\, \Gamma_{0\A}\ ,\nn\\
D_\A&=& e^{-w^\A}\, \fft{\del}{\del z^{\A}} +\ft12 e^{-v}\, w^\A_i\,
\Gamma_{\A i} + \ft12 e^{-w^\B}\, w^\B_\A\, \Gamma_{\A\B}\ ,\\
D_i&=& e^{-v}\, \fft{\del}{\del y^i} +\ft12 e^{-w^\A} \, v_\A
\Gamma_{i\A} +\ft12 e^{-v}\, v_j\, \Gamma_{ij}\ .\nn
\eea
Here, the subscripts on the metric functions denote coordinate-space
derivatives with respect to the indicated coordinates, and the summation
convention applies in the obvious places.  The supercovariant
derivative $\hat D_a$, whose kernel is the space of Killing spinors
associated with unbroken supersymmetries, is then given in $D=11$ by
$\hat D_a = D_a + L_a$, where
\be
L_a = -\ft1{288} F_{bcde}\, \Gamma_a{}^{bcde} + \ft1{36}
F_{abcd}\, \Gamma^{bcd}\ .
\ee

     It is now a straightforward exercise to substitute the
intersecting solutions  into $\hat D_a$, and hence to solve for the
Killing spinors.  For example, for the 
3-intersection solution (\ref{d11m3}), we find that Killing spinors
$\epsilon$ satisfying
$\hat D_a\epsilon=0$ exist provided that
\be
\Gamma_{12}\, \epsilon=\Gamma_{34}\, \epsilon =\Gamma_{56}\,
\epsilon=\Gamma_0\, \epsilon\ ,
\ee
and that $\epsilon= e^{u/2}\, \epsilon_0 = (H_1 H_2 H_3)^{-1/6}\,
\epsilon_0$, where $\epsilon_0$ is a constant.  Thus there are 4
independent solutions, implying that the generalised triple-membrane
intersections preserve $\ft18$ of the supersymmetry.  Note that the
existence of the Killing spinors is assured once the equations of
motion are satisfied.

\section{Conclusions}

        Multi-charge $p$-brane solutions in lower dimensions can be
oxidised to higher dimensions, where they become intersections of
$p$-branes.  To be precise, the direct Kaluza-Klein oxidation gives
rise to intersections of $p$-branes for which the charges are
uniformly distributed over the relative transverse directions.  In
this paper we have generalised these solutions, by allowing the
original harmonic functions that depend only on the overall transverse
space to be functions also of the relative transverse space. We have
argued that these generalised solutions are the more appropriate ones
to consider, since they are the analogues for multi-charge $p$-branes
of the notion of a periodic array of higher-dimensional $p$-branes, as
opposed to a continuum of higher-dimensional charge centres. We found
that such generalised intersections between two isotropic $p$-branes
are not possible.  However, more general solutions do exist, for which
the original independent lower-dimensional harmonic functions become
more general functions that can couple to each other.  Such solutions
can themselves be dimensionally reduced, in the regime where the
distance in the overall transverse space becomes much larger than the
compactification size of the internal space.  These
dimensionally-reduced solutions are then like the standard
lower-dimensional multi-charge solutions.

     These results have the following significance.  Firstly, they
indicate that the lower-dimensional multi-charge solutions, which can
be viewed in the lower dimension as ``bound states at threshold'' (\ie
with zero-binding energy), are not really to be viewed as combinations
of distinct basic objects at all, owing to the interactions in the
higher-dimension that are not observable by a lower-dimensional
observer who is far away from the horizon.  Secondly, since the
lower-dimensional $p$-branes (including RN black holes) are
approximate solutions that are valid for the observer far from the
horizon, then as one approaches the horizon the compactification
dimensions become more and more important, and the solution becomes
essentially higher dimensional.  In this case, the analysis for the
entropy should include the contributions of the non-zero modes.  Of
course there can exist special configurations for which the non-zero
modes are actually vanishing, in which case there will be genuine
no-force conditions between the individual building blocks of the
intersection.  However these solutions, where the $H$ functions in the
higher dimension depend only on the overall transverse coordinates
$\vec y$, are a set of measure zero in the space of all intersections.
The more generic solutions that we have considered in this paper seem
to avoid the problems of the apparent vulnerability of the usual
multi-charge configurations to the ``drifting apart'' of their charge
centres.

\section*{Acknowledgement}

     We are grateful to John Reading and Kelly Stelle for helpful discussions.

\end{document}